\documentclass[aps, prl, twocolumn, notitlepage, groupedaddress, 10pt, floatfix]{revtex4-1}
\usepackage[pdftex,plainpages=false,colorlinks=true,citecolor=blue, linkcolor=blue,urlcolor=blue,filecolor=green,bookmarksopen=true]{hyperref}
\usepackage{natbib}
\usepackage{amsmath}
\usepackage[caption = false]{subfig}
\usepackage{graphicx,epstopdf}
\usepackage[english]{babel}
\usepackage{blindtext}
\usepackage{lipsum}
\usepackage{amsfonts}
\usepackage{amssymb}
\usepackage{bbm}
\usepackage{enumerate}
\usepackage{color}
\usepackage{latexsym}
\usepackage{times,txfonts}
\usepackage{physics}
\usepackage{lipsum}

\newcommand{\us}{\uparrow}
\newcommand{\ds}{\downarrow}

\DeclareSymbolFont{CMletters}{OML}{cmm}{m}{it}
\DeclareMathSymbol{J}{\mathalpha}{CMletters}{`J}
\DeclareMathSymbol{j}{\mathalpha}{CMletters}{`j}
\DeclareMathSymbol{U}{\mathalpha}{CMletters}{`U}

\begin{document}

\title{Giant Magnetoresistance in Boundary-Driven Spin Chains}

\author{Kasper Poulsen}
\email{poulsen@phys.au.dk}
\affiliation{Department of Physics and Astronomy, Aarhus University, Ny munkegade 120, 8000 Aarhus C, Denmark}

\author{Nikolaj T. Zinner}
\email{zinner@phys.au.dk}
\affiliation{Department of Physics and Astronomy, Aarhus University, Ny munkegade 120, 8000 Aarhus C, Denmark}
\affiliation{Aarhus Institute of Advanced Studies, Aarhus University, Høegh-Guldbergs Gade 6B, 8000 Aarhus C, Denmark}

\begin{abstract}
\noindent In solid state physics, giant magnetoresistance is the large change in electrical resistance due to an external magnetic field. Here we show that giant magnetoresistance is possible in a spin chain composed of weakly interacting layers of strongly coupled spins. This is found for all system sizes even down to a minimal system of four spins. The mechanism driving the effect is a mismatch in the energy spectrum resulting in spin excitations being reflected at the boundaries between layers. This mismatch, and thus the current, can be controlled by external magnetic fields resulting in giant magnetoresistance. A simple rule for determining the behavior of the spin transport under the influence of a magnetic field is presented based on the energy levels of the strongly coupled spins.
\end{abstract}

\maketitle

When giant magnetoresistance was first discovered in 1988 \cite{PhysRevLett.61.2472, PhysRevB.39.4828}, it was observed in alternating layers of ferromagnetic and antiferromagnetic materials where an external magnetic field drastically changed the conductive properties of the material. 
Later it has also been observed in ferromagnetic layers separated by isolating nonmagnetic layers \cite{doi:spinvalve_gmr, DIENY1994335, 333913}. 
This discovery led to significant improvements in computer engineering, helping to advance, among others, memory (RAM) \cite{490329}, transistors \cite{PhysRevLett.74.5260}, and sensors \cite{312279}. 
The effect of giant magnetoresistance is largely attributed to electron scattering depending on spin orientation in the aforementioned materials \cite{doi:10.1063/1.122796, 179533}, although recent work found the effect also in one-dimensional Hubbard chains \cite{PhysRevLett.121.020403}.

In the effort of increased miniaturization, traditional electronics has encountered problems due to quantum mechanical effects like tunneling. 
Therefore, many alternative information carriers such as thermal \cite{RevModPhys.84.1045, ROBERTS2011648} and magnetic \cite{Kruglyak_2010, PhysRevLett.91.207901} currents have been proposed. 
Among the most prominent of these is spin transport in boundary-driven spin chains \cite{PhysRevLett.95.057205, PhysRevLett.106.217206, PhysRevLett.107.137201, PhysRevLett.110.047201}. 
Here a linear chain of nearest-neighbor interacting spins are coupled to magnetic reservoirs at both ends, thus inducing a net magnetic transport from one reservoir to the other. 
The current properties are a consequence of the induced steady state which can be engineered into components like diodes \cite{PhysRevE.90.042142, PhysRevLett.120.200603, loss2011}.

Here we show that a class of simple quantum spin systems allow controlled manipulation of the spin current through application of external magnetic fields, i.e., giant magnetoresistance. 
This is done by considering a generic model composed of weakly interacting layers of strongly coupled spins as an analog to the classical phenomenon. 
The groups of strongly coupled spins mimic the action of the ferromagnetic layers which are allowed to interact only weakly with one another, mimicking the insulating layers. 
This results in a mismatch of the energy levels of each group, causing spin excitations to be reflected at the weakly coupled boundary. 
A magnetic field can be applied to align these energy levels, allowing spin excitations to be exchanged, resulting in giant magnetoresistance. 
Our work demonstrates that this coveted and technologically important effect is present in a surprisingly simple quantum system of interacting spins as compared to the condensed-matter materials typically studied. 
Moreover, our work extends the realm of study of giant magnetoresistance to quantum spintronics \cite{Awschalom1174} down to mesoscopic or even few-atom system sizes \cite{fuechsle2012single}. 
As we demonstrate below, the effect can be observed with just a few spins and should be realizable using several of the current platforms used to pursue quantum technology beyond classical electronics. 

\begin{figure*}[t]

\centering
\vspace{0.2cm}

\includegraphics[width=0.9 \linewidth, angle=0]{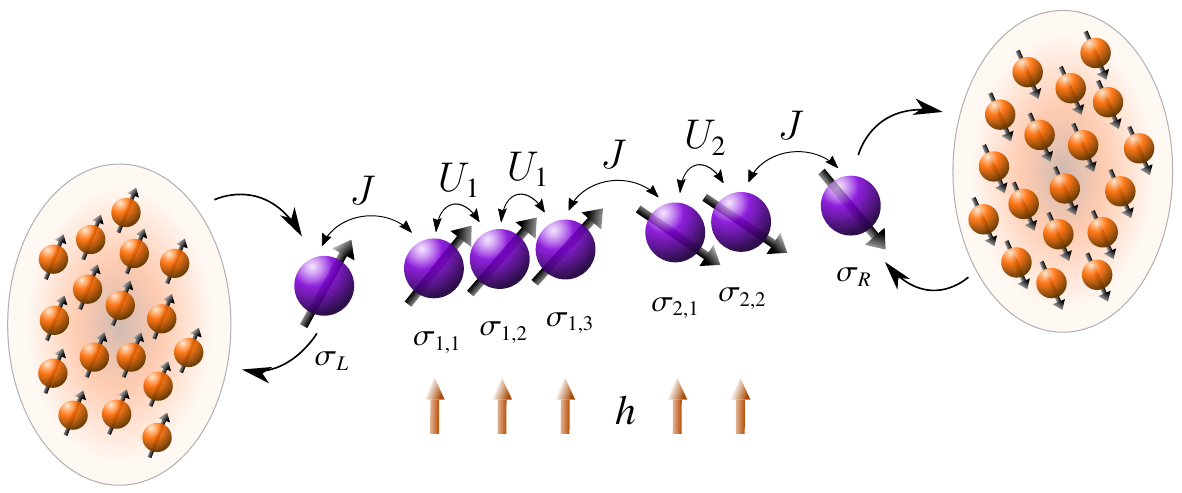}

\caption{%\textbf{Magnetoresistance in a segmented spin chain.} 
Illustration of the model with an example consisting of $N=2$ chains, the first containing $n_1 = 3$ spins and the second $n_2 = 2$ spins. The setup is coupled to spin reservoirs at each end, one with an abundance of spin excitations (left) and one with an abundance of spin excitation holes (right). The exchange coupling between the spins in the first chain is $U_1$, while the exchange between the spins in the second chain is $U_2$. The exchange between the two chains and outer spins is $J$. The numbering is shown below the spins, and the magnetic field is shown with red arrows.}
\label{figure1}
\end{figure*}

{\it Setup.} The general model studied here is a set of $N$ linear spin-1/2 chains where the $i$th chain is composed of $n_i$ spins coupled strongly to each other through the Hamiltonian
\begin{align*}
\hat{H}_0 &= \sum_{i = 1}^{N} \left\{ \sum_{j = 1}^{n_i-1} U_i \left( \hat{\sigma}^x_{i,j} \hat{\sigma}^x_{i,j+1} + \hat{\sigma}_{i,j}^{y} \hat{\sigma}_{i, j+1}^{y} + \Delta_{U_i} \hat{\sigma}_{i,j}^{z} \hat{\sigma}_{i, j+1}^{z} \right) \right. \\ &\hspace{6cm} \left. +\, h \sum_{j = 1}^{n_i} \hat{\sigma}^z_{i,j} \right\}.
\end{align*}
The Pauli matrices for the $j$th spin within the $i$th chain is $\hat{\sigma}^\alpha_{i,j}$ for $\alpha = x,y,z$, and we are using units where $\hbar = 1$. 
The exchange coupling between spins in the $i$th chain is $U_i$, the anisotropy is $\Delta_{U_i}$, and $h$ sets the spin excitation energy for the spins. 
We make these strongly coupled chain segments a part of a larger chain by adding two extra spins labeled $L$ and $R$. 
The extra two spins are described by the Pauli matrices $\hat{\sigma}_{L,R}^\alpha$ for $\alpha = x,y,z$. Finally, we couple these two spins and the strongly interacting chains weakly to each other through the Hamiltonian
\begin{align*}
\hat{H}_{LR} &= J \left( \hat{\sigma}^x_{L} \hat{\sigma}^x_{1,1} + \hat{\sigma}_{L}^{y} \hat{\sigma}_{1,1}^{y} + \Delta_J \hat{\sigma}_{L}^{z} \hat{\sigma}_{1,1}^{z} \right) \\ & \hspace{2cm} + J \left( \hat{\sigma}^x_{N, n_N} \hat{\sigma}^x_{R} + \hat{\sigma}_{N, n_N}^{y} \hat{\sigma}_{R}^{y} + \Delta_J \hat{\sigma}_{N, n_N}^{z} \hat{\sigma}_{R}^{z} \right)\\
\hat{H} =& \, \hat{H}_0 + \hat{H}_{LR} + J \sum_{i=1}^{N-1} \left( \hat{\sigma}^x_{i,n_i} \hat{\sigma}^x_{i+1,1} + \hat{\sigma}_{i,n_i}^{y} \hat{\sigma}_{i+1, 1}^{y} + \Delta_J \hat{\sigma}_{i,n_i}^{z} \hat{\sigma}_{i+1, 1}^{z} \right) 
\end{align*}
where the exchange coupling between chains $J$ must be smaller than the interchain exchanges $J \ll U_i$ and $|\Delta_J|, |\Delta_{U_i}| < 1$. An example of such a setup can be seen in Fig. \ref{figure1}.

To study spin transport in the system, we couple it to spin reservoirs through spin $L$ on the left and spin $R$ on the right; see Fig. \ref{figure1}. The presence of the reservoirs means that we have an open (nonunitary) quantum system that can be described by a density matrix $\hat{\rho}$ and the corresponding Lindblad master equation \cite{Lindblad1976, breuer2002theory}
\begin{align}
\frac{\partial \hat{\rho}}{\partial t} = \mathcal{L}[\hat{\rho}] = -i [\hat{H}, \hat{\rho}] + \mathcal{D}_L [\hat{\rho}] + \mathcal{D}_R [\hat{\rho}].
\label{Lindblad}
\end{align}
Here $[\bullet, \bullet]$ is the commutator, $\mathcal{L}[\hat{\rho}]$ is the Lindblad superoperator, and $\mathcal{D}_{L,R}[\hat{\rho}]$ are dissipative terms describing the action of the baths:
\begin{align*}
\mathcal{D}_{L,R} [\hat{\rho}] &= \gamma \left[ \frac{1 + f_{L,R}}{2}  \left( \hat{\sigma}^+_{L,R} \hat{\rho} \hat{\sigma}^-_{L,R}  - \frac{1}{2} \left\{ \hat{\sigma}^-_{L,R} \hat{\sigma}^+_{L,R} , \hat{\rho} \right\} \right) + \right. \nonumber \\  &\hspace{1.7cm}  \left. \frac{1 - f_{L,R}}{2} \left( \hat{\sigma}^-_{L,R} \hat{\rho} \hat{\sigma}^+_{L,R}  - \frac{1}{2} \left\{ \hat{\sigma}^+_{L,R}\hat{\sigma}^-_{L,R} , \hat{\rho} \right\} \right) \right].
\end{align*}
$\hat{\sigma}^+_{L,R} = (\hat{\sigma}^-_{L,R})^\dag = (\hat{\sigma}^x_{L,R} + i \hat{\sigma}^y_{L,R}) /2$, $\gamma$ is the strength of the interaction with the baths, $f_{L,R}$ determines the nature of the interaction, and $\{\bullet , \bullet \}$ denotes the anticommutator. 
The baths are coupled with strength $\gamma=J$, although we note that smaller values of $\gamma$ induce similar effects. 
The characteristics of these reservoirs are determined by the parameters $f_{L,R}$. 
We will focus on the case where $f = f_L = -f_R$, and unless otherwise stated $f=0.5$. 
One reservoir has an abundance of spin excitations and forces the adjacent spin into a statistical mixture of predominantly up $\ev*{\hat{\sigma}^z_L} = f$, while the other has an abundance of excitation holes and forces the adjacent spin into a statistical mixture of predominantly down $\ev*{\hat{\sigma}^z_R} = -f$.
If $f>0$, on average, spin excitations are created on the left, transported through the chain, and decays on the right, resulting in a current flowing from left to right. However, if $f<0$, the current will tend to flow from right to left. 

The reservoirs will induce currents and generally bring the system out of equilibrium. 
However, after sufficient time, it will reach a steady state (ss), $\frac{\partial \hat{\rho}_{\text{ss}}}{\partial t} = 0$. 
To quantify the spin transport in the steady state, we define the spin current \cite{PhysRevE.90.042142, PhysRevLett.120.200603} as $\mathcal{J} = \tr (\hat{j}_{L}\hat{\rho}_{\text{ss}}) = \tr (\hat{j}_{R}\hat{\rho}_{\text{ss}})$ where $\hat{j}_{L} = 2J \left( \hat{\sigma}^{x}_{L}\hat{\sigma}^{y}_{1,1} - \hat{\sigma}^{y}_{L}\hat{\sigma}^{x}_{1,1}\right)$ and $\hat{j}_{R} = 2J \left( \hat{\sigma}^{x}_{N, n_N}\hat{\sigma}^{y}_{R} - \hat{\sigma}^{y}_{N, n_N}\hat{\sigma}^{x}_{R}\right)$. Note that if the same excitation energy is added to all spins, $\hat{H} \rightarrow \hat{H} + \frac{\omega}{2} \sum_{\alpha} \hat{\sigma}^z_\alpha$, the spin current and the theory presented here would be identical. 
In this case, the spins $L$ and $R$ act as filters, allowing only excitations of frequency $\omega$ to pass. 
This $\omega$ could be an intrinsic excitation energy or a homogeneous magnetic field over the entire system.

\begin{figure}[t]

\centering
\vspace{0.2cm}

\includegraphics[width=1.0 \linewidth, angle=0]{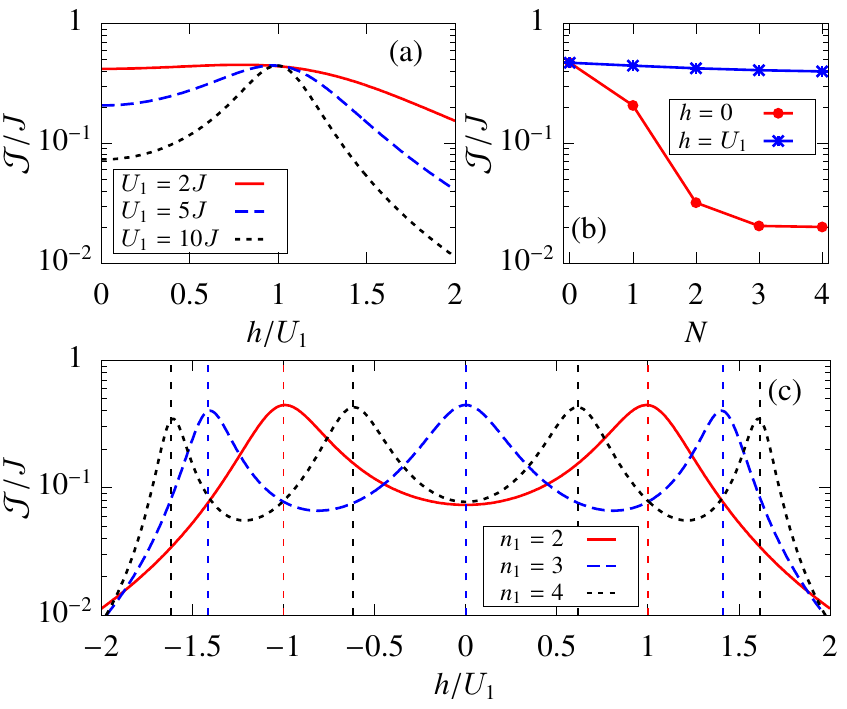}

\caption{%\textbf{Spin current of simple chains.} 
(a) $\mathcal{J}$ as a function of $h/U_1$ for a simple model of only $N=1$ chain consisting of $n_1 = 2$ strongly coupled spins. (b) $\mathcal{J}$ as a function of the number of chains $N$ each consisting of $n_i=2$ strongly coupled spins both on resonance $h=U$ and off resonance $h=0$. For this $U_i = U_1 = 5J$ was used. (c) $\mathcal{J}$ as a function of $h/U_1$ for a single chain ($N=1$) consisting of a different number of strongly coupled spins $n_1$ interacting with an exchange of $U_1 = 10J$. The expected resonances are shown with vertical dashed lines (see the text).}
\label{figure2}
\end{figure}

{\it A single chain.} First, we study the simplest case with $N=1$ chain of $n_1 = 2$ spins coupled strongly to each other with coupling strength $U_1$ and with no anisotropy $\Delta_{U_1}, \Delta_J = 0$. 
This gives a total chain of four spins described by $\hat{\sigma}_L$, $\hat{\sigma}_{1,1}$, $\hat{\sigma}_{1,2}$, and $\hat{\sigma}_{R}$ similar to the example in Fig. \ref{figure1}. 
For this system, an analytical solution can be found for $f=0.5$. The steady state current for $U_1 \gg J$ and $0 \leq h \leq 2U_1$ can be found to be 
\begin{align*}
\mathcal{J}(J, U_1, h) &\approx \frac{h^2 + 17 U_1^2}{\frac{2}{J^2} (h^2 - U_1^2)^2 + \frac{3}{4}(11 h^2 + 43 U_1^2)} J.
\end{align*}
The exact current \footnote{The full current expression for $N=1$, $n_1 = 2$, and $f=1/2$ is given by\\$\mathcal{J}(J, U_1, h) = \frac{2^7 U_1^2( h^2 + 17 U_1^2) + 2^3 17 U_1^2 J^2}{2^8 \frac{U_1^2}{J^2} (h^2 - U_1^2)^2 + 2^5(33 h^2 + 129 U_1^2) U_1^2 + 513 U_1^2 J^2 + 2^4 J^4} J$\\ This was found by putting the master equation \eqref{Lindblad} on a linear form, finding the zero eigenvector, and calculating the current.} is plotted for different values of $U_1$ in Fig.~\ref{figure2}(a). The largest current is obtained for $h = \pm U_1$, where the current is $\mathcal{J}(h=\pm U_1) = \frac{4}{9}J$ and, thus, independent of $U_1$. Furthermore, for no magnetic field $h=0$ the current is $\mathcal{J}(h=0) \sim \frac{17 J^2}{2 U_1^2} J $ to lowest order in $J/U_1$ and, thus, heavily suppressed for large $U_1$. We therefore get giant magnetoresistance even for this minimal model. 
To explain this, we first diagonalize $\hat{H}_0$ to obtain the four states $\ket{\ds \ds}$, $\ket{\Psi_+}$, $\ket{\Psi_-}$, and $\ket{\us \us}$ for spin (1,1) and (1,2) with corresponding energies $E_{\ds\ds} = -2h$, $E_{\Psi_-} = -2U_1$, $E_{\Psi_+} = 2U_1$, and $E_{\us\us} = 2h$, where $\ket{\Psi_\pm} = (\ket{\us \ds} \pm \ket{\ds \us})/\sqrt{2}$. Next, we write the total Hamiltonian $\hat{H}$ in the single excitation basis $\ket{\us \ds \ds \ds}$, $\ket{\ds \!\Psi_+\! \ds}$, $\ket{\ds \!\Psi_-\! \ds}$, and $\ket{\ds \ds \ds \us}$:
\begin{align*}
H = 2 \begin{pmatrix}
-h & \frac{J}{\sqrt{2}} & \frac{J}{\sqrt{2}} & 0\\
\frac{J}{\sqrt{2}} & U_1 & 0 & \frac{J}{\sqrt{2}} \\
\frac{J}{\sqrt{2}} & 0 & -U_1 & -\frac{J}{\sqrt{2}} \\
0 & \frac{J}{\sqrt{2}} & -\frac{J}{\sqrt{2}} & -h
\end{pmatrix}.
\end{align*}
These four states are, therefore, eigenstates with the diagonal being the corresponding eigenenergies of the Hamiltonian to lowest order in $J/U_1$. For a spin excitation created at one end to propagate to the other end, it needs to pass the middle two spins. This is suppressed if the energies of an excitation at either end and an excitation at the middle chain are far from resonance with each other \footnote{This resonance mechanism for controlling spin transport has also been used in quantum spin transistors; see \cite{marchukov2016}.}. 
This also corresponds to an excitation being localized, whereas on resonance, for $h = \pm U_1$, an excitation becomes delocalized over all four spins. Delocalization is known to result in large conductivity within the random dimer model \cite{PhysRevLett.65.88}.
We would therefore expect maxima in the spin current for $h = \pm U_1$ as is also observed in Fig.~\ref{figure2}(a). Remarkably, we see only peaks in the spin current at these two values. Because of the baths, we can expect multiple excitation states to be important. For the simple case of $n_1 = 2$, these can easily be included. An excitation at the left spin can also propagate to the middle two spins through the two transitions
\begin{align*}
\ket{\us \Psi_\pm \ds} \leftrightarrow \ket{\ds \us \us \ds}
\end{align*}
These are likewise at resonance for $h = \pm U$, explaining why only two resonances are observed.

\begin{figure}[t]

\centering
\vspace{0.2cm}

\includegraphics[width=1.0 \linewidth, angle=0]{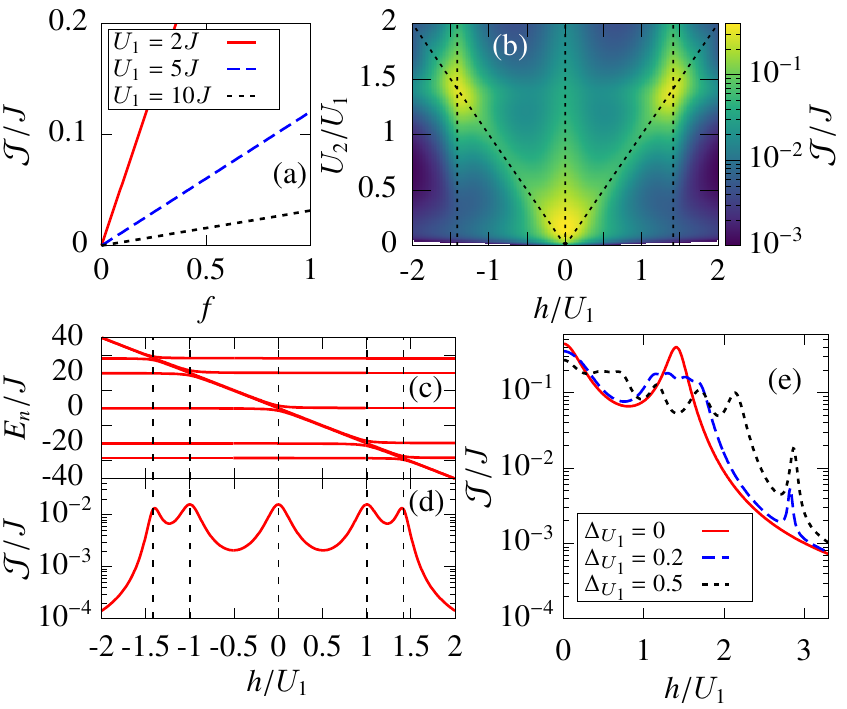}

\caption{%\textbf{Magnetoresistance mechanism.} 
(a) $\mathcal{J}$ as a function of the bath parameter $f$ for the system illustrated in Fig. \ref{figure1} with $U_1 = U_2$ and $h=0$. (b) $\mathcal{J}$ as a function of $h/U_1$ and $U_2/U_1$ with $U_1 = 5J$ as illustrated in Fig. \ref{figure1}. The dashed lines show the expected resonances for both the first chain $h/U_1 = 0,\pm \sqrt{2}$ and the second chain $h/U_2 = \pm 1$. Single excitation spectrum (c) and current $\mathcal{J}$ (d) plotted for the system illustrated in Fig. \ref{figure1} with $U_1 = U_2 = 10J$. The expected resonances are shown with dashed lines for $h/U_1 = 0, \pm 1, \pm \sqrt{2}$. (e) $\mathcal{J}$ as a function of $h/U_1$ for a model of only $N=1$ chain consisting of $n_1 = 3$ strongly interacting spins with hopping $U_1 = 10J$ and anisotropy $\Delta_J = \Delta_{U_1}$.}
\label{figure3}
\end{figure}

{\it Multiple chains.} Keeping $\Delta_{U_i}, \Delta_J = 0$, there are two natural extensions of this, both of which are explored in Figs.~\ref{figure2}(b) and \ref{figure2}(c). 
First, we look at a different number of chains $N$ while keeping $n_i = 2$. We also set all the strong exchange couplings equal $U_i = U_1$. The individual strongly coupled chains diagonalize just as before, and we therefore still expect the strongest current for $h = \pm U_1$. 
The current both off ($h=0$) and at ($h = \pm U_1$) resonance is plotted for a different number of pairs $N$ in Fig. \ref{figure2}(b). 
Off resonance, the spin current is heavily suppressed at first but then levels out for larger $N$, whereas on resonance the current is almost constant. In the limit $N \gg 1$, we likewise expect suppressed current for $h = 0$ and a larger current for $h=U_1$. If instead $n_i$ and $U_i$ are all picked at random, most segments will be off resonant for any $h$ and the current should be suppressed. Therefore, excitations cannot pass between segments similar to Anderson localization. 

Next, we keep $N=1$ and instead vary $n_1$. Following the same process as before, we first diagonalize $\hat{H}_0$. Let $\ket{n}$ be the single excitation state with spin $(1,n)$ flipped. Keeping to this one excitation basis, the Hamiltonian $H_0$ can be written as
\begin{align*}
H_0 = \begin{pmatrix}
(2 - n_1)h & 2U_1 & 0 & \cdots & 0\\
2U_1 & (2 - n_1)h & 2U_1 & \cdots & 0\\
0 & 2U_1 & (2 - n_1)h & \cdots & 0\\
\vdots & \vdots & \vdots & \ddots & \vdots\\
0 & 0 & 0 & \cdots & (2 - n_1)h\\
\end{pmatrix}
\end{align*} 
One can show that the eigenenergies become \cite{losonczi1992eigenvalues, yueh2005eigenvalues}
\begin{align*}
E_k = 4U_1 \cos \left( \frac{\pi k}{n_1 + 1} \right) + (2 - n_1)h, \quad \quad 1 \leq k \leq n_1.
\end{align*}
The corresponding states become eigenstates for $\hat{H}$ to lowest order in $J/U_1$. The states $\ket{\us \ds \ds \! . . .}$ and $\ket{. . . \! \ds \ds \us }$ have energy $-h n_1$ to lowest order. Therefore, an excitation at the ends is resonant with an excitation in the chain when $h = 2U_1 \cos \left( \frac{\pi k}{n_1 + 1} \right)$ for $1 \leq k \leq n_1$. 
For $n_1 = 2$, this reduces to $h/U_1 = \pm 1$ as expected. As a few more examples, we get $h/U_1 = 0,\pm \sqrt{2}$ for $n_1 = 3$ and the four solutions $h/U_1 = \pm (\sqrt{5} \pm 1)/4$ for $n_1 = 4$. 
The spin current is plotted for these three examples in Fig.~\ref{figure2}(c), and the resonances found above are plotted with vertical dashed lines. 
The maxima in the current occur at the resonances as expected, while the current is suppressed away from them. 
Once again, we observe maxima only at the values found above even though our analysis only includes single excitation states. 
Like for the simple case, the entire Hilbert space for the strongly interacting spins could be included, but this becomes increasingly difficult as $n_1$ increases. 
We have checked numerically that all resonances occur at $h = 2U_1 \cos \left( \frac{\pi k}{n_1 + 1} \right)$ for $1 \leq k \leq n_1$ for all cases $n_1 \leq 15$; see Supplemental Material for more details. 
Especially for $|h| > 2U_1 \cos \left( \frac{\pi}{n_1 + 1} \right)$ is the current heavily suppressed. In the thermodynamic limit $n_1 \gg 1$, the single excitation spectrum for the strongly interacting chain approaches a continuum in the interval $-4U_1 < E_k < 4U_1$, and an appreciable current is expected for $-2U_1 < h < 2U_1$, while a hard dropoff should occur for $|h| > 2U_1$.

In the more general case, we look at $N=2$ chains consisting of $n_1 = 3$ and $n_2 = 2$ strongly coupled spins, respectively, as seen in Fig. \ref{figure1}. 
At first, we keep $U_1 \neq U_2$. The first chain will then be at resonance with the ends for $h/U_1 = 0, \pm \sqrt{2}$, while the second chain will be at resonance with the ends for $h/U_2 = \pm 1$. 
However, only when both the chains individually are at resonance with each other so that a spin excitation can propagate between them do we expect the largest current. This is the case when both of the above conditions are upheld or rather when $U_2 = \pm \sqrt{2} U_1$ or $U_2 \sim 0$. 
To see that this is true, we plot the current as a function of both $U_2/U_1$ and $h/U_1$ in Fig.~\ref{figure3}(b) with the expected resonances plotted as dashed lines. Here we see that lines of high current run along the expected lines and that the current is extra large when the resonances meet. 
To illustrate the role of the single excitation spectrum explored above, we set $U_1 = U_2 = U$ and plot both the current and the single excitation spectrum as a function of $h/U$ in Figs. \ref{figure3}(c) and \ref{figure3}(d), respectively. 
Again, we plot the expected resonances with dashed lines. The two eigenenergies that are linearly dependent on $h/U$ corresponds to eigenstates that are close to $\ket{\us \ds ... \ds}$ and $\ket{\ds ... \ds \us}$, whereas the others are close to eigenstates that correspond to a spin excitation within the strongly coupled chains. 
Here it is clearly seen that, when the energies of the states describing excitations at the ends cross the energy of the states with excitations within the chains, a higher current is observed.
Hence, we see that the giant magnetoresistance is attributed to a set of resonance conditions that can be predicted for particular setups. This leads to several generalizations. First, for a large number of strongly interacting chains $N\gg 1$ with a random number of spins $n_i$, the excitation will be scattered at most boundaries, thus resulting in poor conductivity.
Second, if the spins $L$ and $R$ are substituted for general systems, a resonance will be observed when the frequencies of these systems are resonant with the neighboring strongly interacting chain.

We address the question of sensitivity to the nature of the bath parameter 
$f$ in Fig. \ref{figure3}(a). Here it is seen that the current depends linearly on $f$, and, therefore, the effects studied above will be present for any $f>0$.

{\it Including Z-kobling.} Finally, we include anisotropy for a model of only $N=1$ chain of $n_1 = 3$ strongly interacting spins in Fig. \ref{figure3}(e). The addition of anisotropy has two main effects. First, the spectrum is perturbed, moving and splitting up the resonances.  Second, a new peak appears for $h/U_1 \sim 2 \sqrt{2}$ due to the transition 
$\ket*{\us \Lambda_{\us \ds \ds} \ds} \rightarrow \ket*{\ds \Lambda_{\us \us \ds} \ds} \rightarrow \ket*{\ds \Lambda_{\us \ds \ds} \us}$, where
\begin{align*}
\ket*{\Lambda_{\us \ds \ds}} &= \frac{1}{2} \left( \ket{\us \ds \ds} - \sqrt{2} \ket{\ds \us \ds} + \ket{\ds \ds \us} \right) + O(\Delta_{U_1})\\
\ket*{\Lambda_{\us \us \ds}} &= \frac{1}{2} \left( \ket{\us \us \ds} + \sqrt{2} \ket{\us \ds \us} + \ket{\ds \us \us} \right) + O(\Delta_{U_1})
\end{align*}
to lowest order in $\Delta_{U_1}$. The matrix element for the first transition is
\begin{align*}
\mel{\ds \Lambda_{\us \us \ds} \ds}{\hat{H}}{\us \Lambda_{\us \ds \ds} \ds} = -\frac{\Delta_{U_1}}{4} + O(\Delta_{U_1}^3).
\end{align*}
For this, higher order terms of $\ket*{\Lambda_{\us \ds \ds}}$ and $\ket*{\Lambda_{\us \us \ds}}$ were included. The matrix element is zero for $\Delta_{U_1}= 0$, explaining why the resonance is absent for this case. The transition is at resonance for $h/U_1 = 2\sqrt{2} \left[ 1+\frac{\Delta_{U_1}^2}{16} + O(\Delta_{U_1}^4) \right]$.
The effects of including anisotropy is further studied in Supplemental Material.

{\it Conclusion.} We have shown how a system of weakly interacting layers of strongly coupled spins exhibits the defining quality of giant magnetoresistance; i.e., we can control the spin current in the chain by applying external magnetic fields. This is caused by reflection of spin excitations at the boundaries between the strongly coupled regions when a mismatch in the energy levels is present. We show that the effect is present even in the simplest case of four spins by obtaining an analytical expression for the spin current, and we propose a method for finding large current resonances in a general chain. This provides a 
simple picture for understanding and predicting giant magnetoresistance in spin chains. 
The spin model studied here is generic with many implementation possibilities including neutral atoms in optical lattices \cite{Simon2011, PhysRevLett.93.250405}, phosphourus-doped silicon surfaces \cite{Awschalom1174, Morton2011}, or superconducting circuits \cite{devoret2013}. A possible implementation using superconducting circuits is proposed in Supplemental Material.

\begin{acknowledgments}
The authors acknowledge funding from The Independent Research Fund Denmark DFF-FNU.
\end{acknowledgments}

\bibliography{bibliography}

\onecolumngrid
{\color{white} 0}

\newpage

\renewcommand{\thesubsection}{Appendix A\arabic{subsection}}
\appendix

\section*{ \huge{ S\lowercase{upplemental} M\lowercase{aterial} }}

\subsection{Effects of anisotropy}

\begin{figure}[t]

\centering
\vspace{0.2cm}

\includegraphics[width=0.6 \linewidth, angle=0]{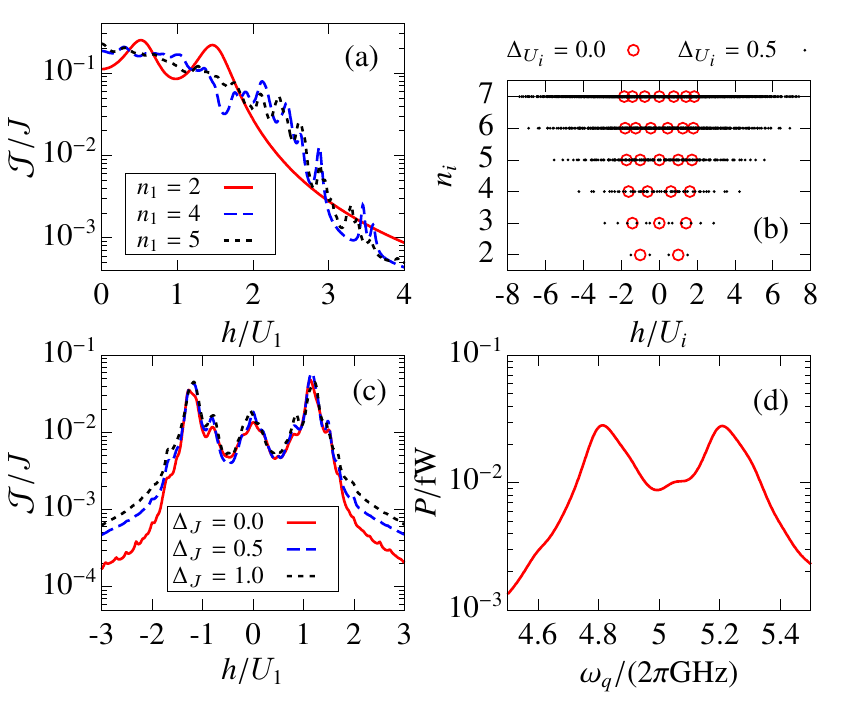}

\caption{(a) $\mathcal{J}$ as a function of $h/U_1$ for $N=1$ chain of $n_1$ strongly coupled spins where $\Delta_{U_1} = \Delta_J = 0.5$ and $U_1 = 10J$. (b) Resonance $h$-values for different chain lengths $n_i$ with and without anisotropy. (c) $\mathcal{J}$ as a function of $h/U_1$ for the system illustrated in Fig. \ref{figure1} in the main article ($N=2$, $n_1= 3$ and $n_2 = 2$) where $U_1 = U_2 = 10J$ and $\Delta_{U_1} = 0.2$. (d) Power $P$ as a function of transmon energy $\omega_q$ for the superconducting circuit in Fig. \ref{figure5} and the parameter in \eqref{param}.}
\label{figure4}
\end{figure}

In the main article, we observed that adding a Z-coupling has two main effects. First, degeneracies in the spectrum of the strongly interacting spins are lifted thus perturbing and splitting up the resonances. 
Second, previously forbidden transitions become allowed due to either the extra term in the Hamiltonian or the energy eigenstates for the strongly interacting spins changing. 
To expand on this point, we plot the current for $N=1$ chain of strongly interacting spins as a function of $h$ for different $n_1$ in Fig.~\ref{figure4}(a). 
For the case of $n_1 = 2$ strongly interacting spins, there are four different ways for an excitation to travel from the left spin to the two middle spins. These are
\begin{align*}
\ket{\us \ds \ds \ds} & \rightarrow \ket{\ds \Psi_- \ds} \\
\ket{\us \ds \ds \ds} & \rightarrow \ket{\ds \Psi_+ \ds} \\
\ket{\us \Psi_- \ds} & \rightarrow \ket{\ds \us \us \ds} \\
\ket{\us \Psi_+ \ds} & \rightarrow \ket{\ds \us \us \ds}.
\end{align*}
For $\Delta_{U_1} = 0$, these four transitions results in two resonances. The first and fourth transition obey energy conservation for $h=U$, while the second and third obeys energy conservation for $h=-U$. For $\Delta_{U_1} \neq 0$, the four transitions obey energy conservation for different $h$ values thus resulting in four resonances. The second effect only occurs for $n_1 \geq 3$. The case of $n_1 = 3$ is studied in the main article. For $n_1 \geq 4$ the combination of the two effects makes it difficult to identify each resonance as can be seen in Fig. \ref{figure4}(a).

To further understand the cases $n_1 \geq 4$, we plot the resonant $h$-values for different lengths of strongly interacting spins with and without anisotropy in Fig.~\ref{figure4}(b). 
These resonances were found numerically by looping over all pairs of eigenstates $\ket{E_\alpha}$ and $\ket{E_{\alpha'}}$ for $\hat{H}_0$ where $N=1$. 
Only transitions allowed by $\hat{H}$ to first order are included, that is $\mel*{\us\! E_\alpha \! \ds}{\hat{H}}{\ds\! E_{\alpha'}\! \ds} \neq 0$. 
Without anisotropy the resonances are exactly as expected from the analysis in the main article. Note that this numerical result include the entire Hilbert space, however, it does not take into account how strongly $\hat{H}$ couples the corresponding eigenstates. 
As mentioned in the main article, the calculation was also done for all $n_1 \leq 15$ without anisotropy to check that the resonances are given by $h = 2U_1 \cos \left( \frac{\pi k}{n_1 + 1} \right)$ for $1 \leq k \leq n_1$. 
However, only resonances for $n_1 \leq 7$ are plotted here. With anisotropy we see the expected resonances for $n_1 =2,3$, and we can even identify some resonances in Fig.~\ref{figure4}(a) from the data in Fig.~\ref{figure4}(b) for $n_1 = 4,5$. It is also seen that the exponential increase in the size of the Hilbert space as a function of $n_1$ results in an increasing number of resonances. These resonances will overlap and make the giant magnetoresistance effect smaller. This can already be seen for $n_1 = 5$.

The effect of $\Delta_J$ can be seen in Fig.~\ref{figure4}(c) where the current is plotted for a model of two chains of $n_1 = 3$ and $n_2 = 2$ strongly interactive spins respectively. The main effect is seen for larger $|h|$ where the current is larger with anisotropy. The general behavior is not changed.

\section{Experimental setup}

\begin{figure}[t]

\centering
\vspace{0.2cm}

\includegraphics[width=0.7 \linewidth, angle=0]{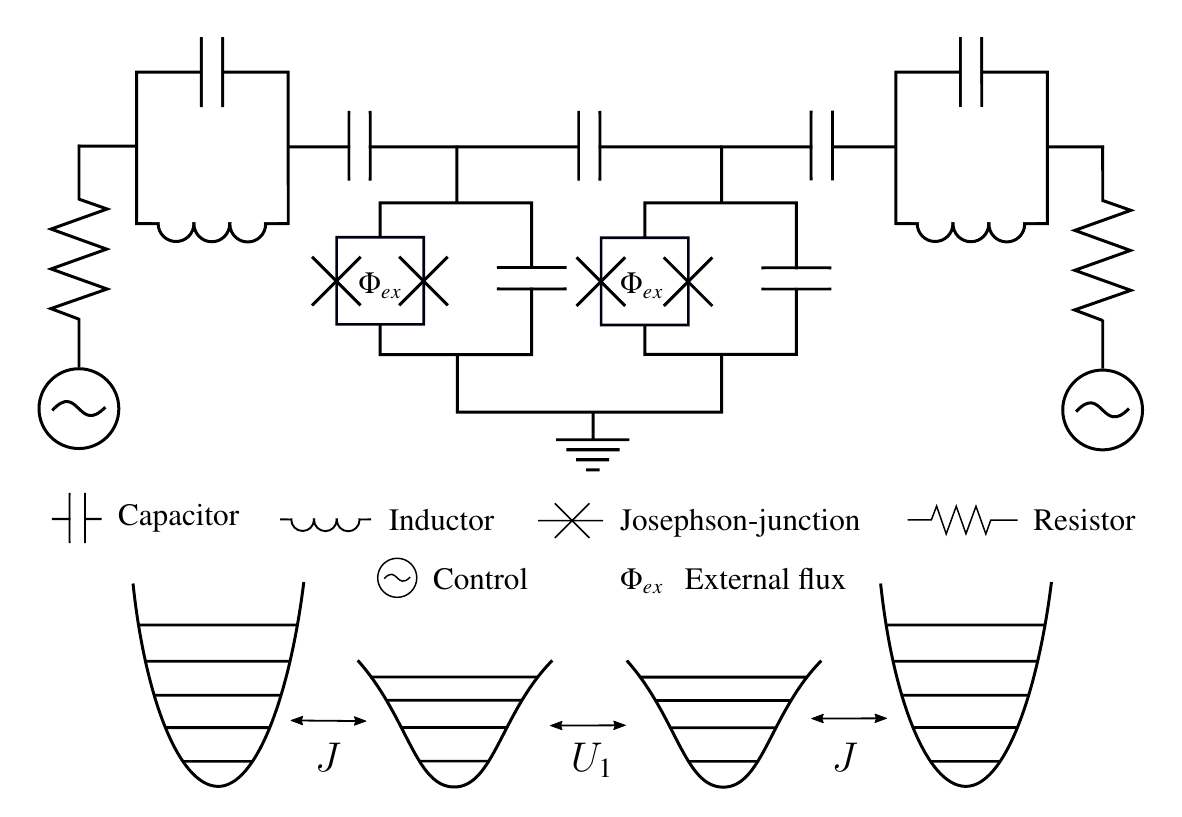}

\caption{(Top) Superconducting circuit implementation of the minimal model of $N=1$ chain of $n_1 = 2$ strongly interacting spins. (Bottom) Schematic representation of the circuit with two harmonic oscillators and two transmons. 
The harmonic oscillators are drawn as a harmonic potential with equally spaced energy levels while the transmons are drawn as the bottom of a cosine potential \citep{doi:10.1063/1.5089550}.}
\label{figure5}
\end{figure}

One possible experimental setup of the minimal model in superconducting circuits can be seen in Fig.~\ref{figure5}.
Analyzing the circuit in the figure in the proper regime, one can get a Hamilton describing two harmonic oscillator connected by two transmon qubits. 
In an experiment the harmonic oscillators would consist of two superconducting resonators with capacitive coupling to the two transmons \cite{Senior2020}. 
The correlation functions of the resonators decay due to the resistor thus acting like a bath. 
If these correlation functions decay fast enough, one can see the entire resonator as a bath with a spectral density proportional to a Lorenzian type function \cite{PhysRevB.101.184510}. 
However, to make the analogy to the model from the main article stronger we will model the resonator as a harmonic oscillator coupled to a thermal bath with strength $\gamma$. 
This model reduces to a Lorenzian spectral density for $\gamma \gg J$. Performing second quantization and neglecting next-nearest neighbor interaction, the Hamiltonian can be written
\begin{align*}
\hat{H} &= \omega\, (\hat{n}_L + \hat{n}_R) + \sum_{i = 1}^2 \omega_q\, \hat{n}_{1,i} -  \frac{\delta \omega_q}{2} \, \hat{n}_{1,i} (\hat{n}_{1,i} - 1) \\
& \hspace{4.cm}+ 2 J \left( \hat{a}_L^\dag \hat{a}_{1,1} + \hat{a}_L \hat{a}_{1,1}^\dag \right) + 2 U_1 \left( \hat{a}_{1,1}^\dag \hat{a}_{1,2} + \hat{a}_{1,1} \hat{a}_{1,2}^\dag \right) + 2 J \left( \hat{a}_{1,2}^\dag \hat{a}_R + \hat{a}_{1,2} \hat{a}_R^\dag \right)
\end{align*}
where $\hat{a}_L$ ($\hat{a}_R$) is the annihilation operator for the left (right) harmonic oscillator \cite{doi:10.1063/1.5089550}. 
Since the system is coupled to baths, we include higher excited levels of the transmons and write the corresponding annihilation operator $\hat{a}_{1,1}$ and $\hat{a}_{1,2}$. 
The corresponding number operator is $n_{1,i} = \hat{a}_{1,i}^\dag \hat{a}_{1,i}$. $\omega$ denotes the excitation energy of the harmonic oscillator and $\omega_q$ of the transmons. $\delta \omega_q$ sets the anharmonicity of the transmons. 
Analogous to the original model the transmon energy is $\omega_q = \omega + 2h$. The transmon energy can be tuned through the external flux $\Phi_{ex}$. The time-evolution of the density matrix is still determined by the master equation (1), in the main article, where
\begin{align*}
\mathcal{D}_{L,R} [\hat{\rho}] &= \gamma \left[ n_{L,R}  \left( \hat{a}^\dag_{L,R} \hat{\rho} \hat{a}_{L,R}  - \frac{1}{2} \left\{ \hat{a}_{L,R} \hat{a}^\dag_{L,R} , \hat{\rho} \right\} \right) + (n_{L,R} + 1)  \left( \hat{a}_{L,R} \hat{\rho} \hat{a}^\dag_{L,R}  - \frac{1}{2} \left\{ \hat{a}^\dag_{L,R} \hat{a}_{L,R} , \hat{\rho} \right\} \right) \right]
\end{align*}
where
$$n_{L,R} = \left( e^{\omega/T_{L,R}} -1 \right)^{-1}.$$
$T_{L,R}$ is the temperature of the respective bath and $k_B = 1$. The spin current would here be measured as an energy current, $\mathcal{K} = \omega \mathcal{J}$, measured in watts. Some typical values for superconducting circuits are
\begin{align}
J/2\pi &= 10\mathrm{MHz} & U/2\pi &= 100 \mathrm{MHz} & \omega /2\pi &= 5\mathrm{GHz} \nonumber\\ 
\gamma &= 600\mathrm{MHz} & T_H &= 240\mathrm{mK} & T_C &= 50\mathrm{mK} \label{param} \\
& & \delta \omega / \omega_q &= 0.02 & & \nonumber 
\end{align}
The power for these values is plotted in Fig.~\ref{figure4} (d) as a function of $\omega_q$. For the simulation the lowest five states of the harmonic oscillators and the lowest three states of the transmons were included. The resonances are expected at $\omega_q/2\pi =  5\mathrm{GHz} \pm 0.2\mathrm{GHz} $, which is also observed. However, the resonances are not symmetric anymore due to the effects of the higher excited levels of the transmons.

\end{document}